\documentclass[%
reprint,
superscriptaddress,
amsmath,amssymb,
prb,
]{revtex4-2}

\usepackage{graphicx}
\usepackage{dcolumn}
\usepackage{bm}
\usepackage{setspace}

\usepackage[caption=false]{subfig}

\usepackage{xcolor}
\usepackage{upgreek}
\usepackage{hyperref}
\hypersetup{
	colorlinks = true,
	linkcolor  = {red},
	citecolor  = {red},
	urlcolor = {blue}
}

\usepackage[utf8]{inputenc}
\usepackage[T1]{fontenc}
\usepackage{mathptmx} 

\begin{document}
\setstretch{1.25}
\title{Tailoring the switching efficiency of magnetic tunnel junctions by the fieldlike spin-orbit torque}

\author{Viola Krizakova}
    \email{viola.krizakova@mat.ethz.ch}
     \affiliation{Department of Materials, ETH Zurich, 8093 Zurich, Switzerland}
\author{Marco Hoffmann}
     \affiliation{Department of Materials, ETH Zurich, 8093 Zurich, Switzerland}
\author{Vaishnavi Kateel}%
    \affiliation{imec, Kapeldreef 75, 3001 Leuven, Belgium}%
\author{Siddharth Rao}%
    \affiliation{imec, Kapeldreef 75, 3001 Leuven, Belgium}%
 \author{Sebastien Couet}%
    \affiliation{imec, Kapeldreef 75, 3001 Leuven, Belgium}%
\author{Gouri Sankar Kar}%
    \affiliation{imec, Kapeldreef 75, 3001 Leuven, Belgium}%
\author{Kevin Garello}%
    \affiliation{Université Grenoble Alpes, CEA, CNRS, Grenoble INP, SPINTEC, 38054 Grenoble, France}%
\author{Pietro Gambardella}
   \email{pietro.gambardella@mat.ethz.ch}
    \affiliation{Department of Materials, ETH Zurich, 8093 Zurich, Switzerland}

\begin{abstract}
Current-induced spin-orbit torques provide a versatile tool for switching magnetic devices. 
In perpendicular magnets, the dampinglike component of the torque is the main driver of magnetization reversal. The degree to which the fieldlike torque assists the switching is a matter of debate. Here we study the switching of magnetic tunnel junctions with a CoFeB free layer and either W or Ta underlayers, which have a ratio of fieldlike to dampinglike torque of 0.3 and 1, respectively. We show that the fieldlike torque can either assist or hinder the switching of CoFeB when the static in-plane magnetic field required to define the polarity of spin-orbit torque switching has a component transverse to the current. In particular, the non-collinear alignment of the field and current can be exploited to increase the switching efficiency and reliability compared to the standard collinear alignment. By probing individual switching events in real-time, we also show that the combination of transverse magnetic field and fieldlike torque can accelerate or decelerate the reversal onset. We validate our observations using micromagnetic simulations and extrapolate the results to materials with different torque ratios. Finally, we propose device geometries that leverage the fieldlike torque for density increase in memory applications and synaptic weight generation.
\end{abstract}

\maketitle


\section{Introduction}
\vspace{-6pt}

Current-induced spin-orbit torques (SOT) \cite{Manchon2019} offer an efficient and scalable way to control the magnetization of spintronic devices \cite{Miron2011}, including magnetic tunnel junctions (MTJ) \cite{Liu2012,Cubukcu2014,Grimaldi2020,Krizakova2021,Zhang2021a} domain wall racetracks \cite{Miron2011NatMat,Emori2013a,Yang2015,Raymenants2021a} and logic gates \cite{Luo2021a,Alamdar2021,Bhowmik2014,Baek2018a}. For its relevance in memory and computing applications, SOT switching has undergone much progress in terms of reliability, operation speed, energy efficiency, as well as realizing zero-external-field switching in systems with perpendicular magnetization \cite{Krizakova2022jmmm}.
In these systems, which offer the best scaling prospects in terms of device integration, a static magnetic field along the current direction is required to break the SOT symmetry and determine the switching polarity \cite{Miron2011}. The SOT can be decomposed into two orthogonal components, the longitudinal dampinglike torque (DLT) and the transverse fieldlike torque (FLT) \cite{Manchon2019,Garello2013,Kim2013,Avci2014}. Most work on switching, however, concentrates on the DLT, for it is known to drive the magnetization reversal when assisted by the longitudinal field \cite{Manchon2019,Miron2011,Liu2012,Lee2013,Finocchio2013}, whereas the FLT is often disregarded. Similarly to a magnetic field of fixed orientation, however, the FLT has different effects on the reversal dynamics. In the macrospin picture, it promotes the precession of the magnetization about the direction perpendicular to the current flow \cite{Legrand2015}, whereas, in the case of incoherent magnetization reversal, it lowers the energy barrier for domain nucleation \cite{Miron2010} and can favor the propagation of domain walls \cite{Baumgartner2017}. The FLT can therefore be probed by studying the switching in different current–field configurations.

An in-plane magnetic field applied perpendicular to the current was found to promote (hinder) the nucleation of magnetic domains \cite{Miron2010} and to reduce (increase) the switching threshold when applied along (against) the effective field of the FLT \cite{Baumgartner2017}. 
In samples with strong FLT, varying the direction of the assisting magnetic field in the plane was found to change the threshold current for SOT switching \cite{Fan2019} and the onset of the backward switching \cite{Lee2018a}, which could be exploited to realize unipolar switching. These results show that the FLT acts as an internal effective field that superposes to the external one. Simulations also suggest that the FLT can enable field-free switching, supposing materials with a specific FLT-to-DLT ratio ($\beta$) are employed \cite{Legrand2015,Hassan2019,Wang2021}.
On the other hand, the FLT can increase the switching threshold by tilting the magnetization into the plane and induce precessional dynamics, which impairs the switching reliability \cite{Jiang2020}. The increased susceptibility to switching errors, observed in strong-FLT samples, was attributed to the domain-wall reflection from the edges of the magnetic structure \cite{Yoon2017,Lee2018a}. Thus, whereas magnetization reversal is relatively well understood in SOT materials in which the DLT dominates over the FLT, such as Pt \cite{Garello2013,Emori2013a,Nguyen2016} and W \cite{Pai2014,Garello2018,Sethu2021}, strong-FLT materials, such as Ta \cite{Garello2013,Kim2013,Avci2014,Torrejon2014}, Hf \cite{Pai2014,Torrejon2014,Ramaswamy2016,Ou2016} and topological insulators \cite{Mellnik2014,Binda2021,Bonell2020}, offer additional opportunities to tune the switching efficiency. Experimental work addressing the FLT, however, has only focused on the switching of relatively large ($\upmu$m-scale) structures, and used either Pt \cite{Miron2010,Baumgartner2017} or Ta \cite{Yoon2017,Lee2018a,Fan2019} as the SOT source. The role of the FLT in the reversal dynamics and its effect on the threshold current of nanoscale devices such as MTJs are not known.

In this paper, we investigate the influence of the FLT and in-plane external field on the switching of nanoscale MTJ devices based on W/CoFeB and Ta/CoFeB heavy metal/ferromagnetic layers. Our results elucidate the impact of the FLT on the switching dynamics, reliability, and threshold conditions for both low- and strong-FLT systems, showing that it can have advantages for practical applications. By measuring individual switching events in the time domain, we find that the FLT directly affects the energy barrier for reversal and accelerates or decelerates the switching onset. We also show that $\beta$ can be estimated at device level from the switching measurements in presence of a transverse field. Using micromagnetic simulations, we further elucidate the effects of the FLT and in-plane external field and extrapolate the results to materials with different FLT strengths.
Finally, we propose device geometries that can leverage the interplay of the external field and FLT for either high-density memory applications or the generation of synaptic weights.

\section{\label{sec:experiment}Experiment}
\vspace{-6pt}

\begin{figure}
\includegraphics[width=85mm]{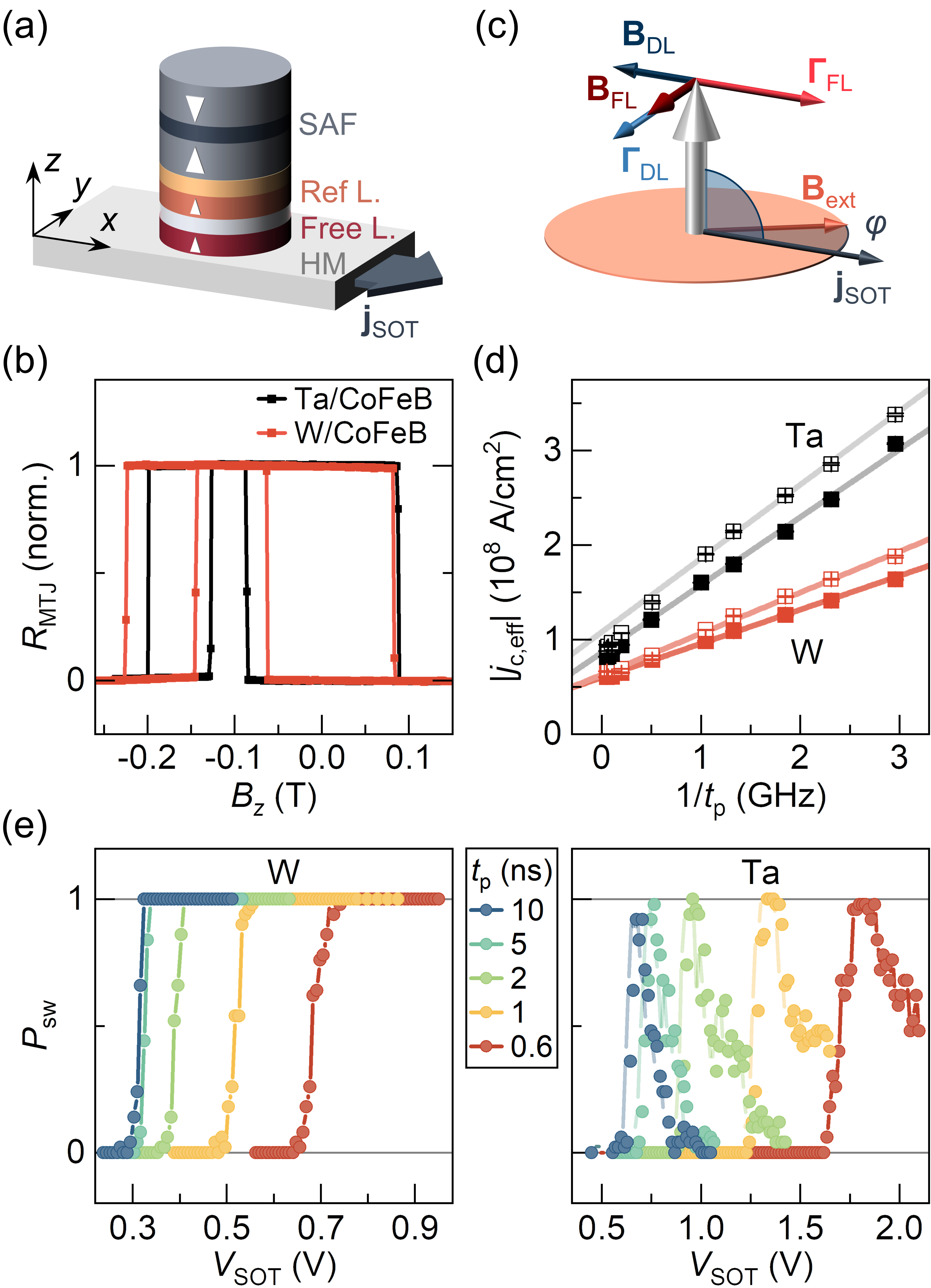}
\subfloat{\label{fig1:a}}%
\subfloat{\label{fig1:b}}%
\subfloat{\label{fig1:c}}%
\subfloat{\label{fig1:d}}%
\subfloat{\label{fig1:e}}%
\caption{\label{fig1:Experiment} (a) Schematics of the sample geometry. (b) Comparison of the normalized hysteresis loops of MTJ with W and Ta underlayers. (c) Schematics of the dampinglike and fieldlike SOT $\Gamma_\text{DL,FL}$ and the corresponding effective fields $B_\text{DL,FL}$ induced by $j_{\text{SOT}}$. (d) Comparison of the effective critical switching currents $j_{\text{c,eff}}$ for different pulse widths obtained for an applied in-plane field $B_x = 32$\,mT. Open (full) symbols indicate switching to the up (down) state. Lines are linear fits to the data in the intrinsic regime ($1/t_\text{p} \geq 1$\,GHz). (e) Probability of switching to the up state as a function of the SOT bias out of 50 trials.}
\end{figure}

We use MTJ devices patterned into circular pillars with the diameter of 80\,nm [Fig.\,\ref{fig1:a}] and grown on top of a heavy metal (HM) current injection track. The device structure is HM/CoFeB/MgO/CoFeB(1.1)/SAF(10.5), where the numbers in parentheses indicate the thickness in nanometers. The synthetic antiferromagnet (SAF) is used to pin the upper CoFeB layer upward. In the study, we compare MTJs comprising $\beta$-W(3.5)/CoFeB(0.9) and Ta(5)/CoFeB(1). The resistivity of the W, Ta, and CoFeB layers is estimated to be 160\,$\upmu\Omega$\,cm, 210\,$\upmu\Omega$\,cm, and 120\,$\upmu\Omega$\,cm, respectively. The efficiency of the DLT\,(FLT) obtained from harmonic Hall measurements \cite{Garello2013} are $\xi_\text{DL(FL)} = -0.33\pm0.03 (-0.10\pm0.02)$ for the W-based \cite{Sethu2021} and $\xi_\text{DL(FL)} = -0.11\pm0.01 (-0.11\pm0.03)$ for the Ta-based samples.
In both types of samples, the free layer has an easy axis along $z$ and its magnetization can be reversed between up and down states without any stable intermediate levels [Fig.\,\ref{fig1:b}]. Moreover, the SAF structure creates a dipolar field, $|B_\text{SAF}| \leq 10$\,mT along $-z$ that favors the up-to-down reversal of the free layer in all samples.

In the experiment, a positive $V_\text{SOT}$ applied across the SOT track induces a current $j_\text{SOT}$ along the $x$ direction. This current generates dampinglike and fieldlike SOT on the bottom CoFeB (free layer) magnetization, as shown in Fig.\,\ref{fig1:c}. The magnetic field $B_\text{ext}$ is applied in the $xy$ plane along the direction given by the angle $\varphi$ with respect to $x$.
The final state after the SOT pulse injection is read by applying a small oscillating voltage (10\,mV) on the MTJ. Additionally, the free layer magnetization can be probed during the SOT pulse, in order to perform the time-resolved measurements discussed in Sec.\,\ref{sec:TR}. To do that, a small current shunt across the pillar ($<1.3$\,MA/cm$^2$) is used to read the real-time resistance change of the junction on an oscilloscope \cite{Grimaldi2020}. This resistive change corresponds to the change of the magnetization direction.

\section{SOT switching for collinear alignment of current and field}
\vspace{-6pt}

First, we compare the SOT switching in both types of MTJ for collinear alignment of current and in-plane magnetic filed ($\varphi = 0^\circ$). We applied a 32\,mT field along $x$ and measured the probability of switching $P_\text{sw}$ of the free layer upon repeated pulsing with different amplitude $V_\text{SOT}$. After each pulse, we read the MTJ state and reset it to the initial state afterwards. We repeated this procedure for different pulse widths $t_\text{p}$. Figure\,\ref{fig1:d} shows the effective critical current $j_\text{c,eff}$, defined as the current density in and below the free layer for which $P_\text{sw} = 0.5$ \cite{Wu2021}. In the short-pulse limit, the critical current scales as $j_\text{c,eff} = j_\text{c0} + q/t_\text{p}$, where $j_\text{c0}$ is the intrinsic critical current and $q$ is the effective charge parameter that determines the rate at which angular momentum is transferred to the free layer \cite{Bedau2010,Liu2014a,Garello2014,Krizakova2020}. From the linear fit to the data, we obtain $j_\text{c0} = 62\pm 2$\,MA/cm$^{2}$ and $q = 396\pm 38$\,C/m$^{2}$ for W, and $ j_\text{c0} = 97\pm 15$\,MA/cm$^{2}$ and $q = 748\pm 44$\,C/m$^{2}$ for Ta. The small difference in $j_\text{c0}$ can be attributed to difference in the thermal stability and heat dissipation in the system, whereas and the factor of $\approx 2$ between the $q$ in W and Ta reflects mainly the relative difference between the DLT efficiencies of these metals \cite{Manchon2019,Cao2020}.

Figure\,\ref{fig1:e} shows $P_\text{sw}$ for down-to-up switching, which is significantly different for W and Ta. In the case of W, the switching to the opposite state remains reliable after reaching $P_\text{sw} = 1$. In the case of Ta, however, the increase of $P_\text{sw}$ with $V_\text{SOT}$ toward 1 is followed by a gradual increase of error rate, leaving only a very limited interval of voltages suitable for reliable operation. This behavior is not unexpected in Ta/CoFeB, as similar observation have been reported earlier \cite{Yoon2017,Lee2018a} and attributed to the FLT. However, contrary to Ref.~\onlinecite{Lee2018a}, in which Hall cross samples with a single magnetic layer were used, in the MTJ devices we only observe the onset of large switching errors for the one of the two switching directions that is opposed by $B_\text{SAF}$.

\section{SOT switching for noncollinear alignment of current and field}
\vspace{-6pt}

\begin{figure}[bt]
\includegraphics[width=85mm]{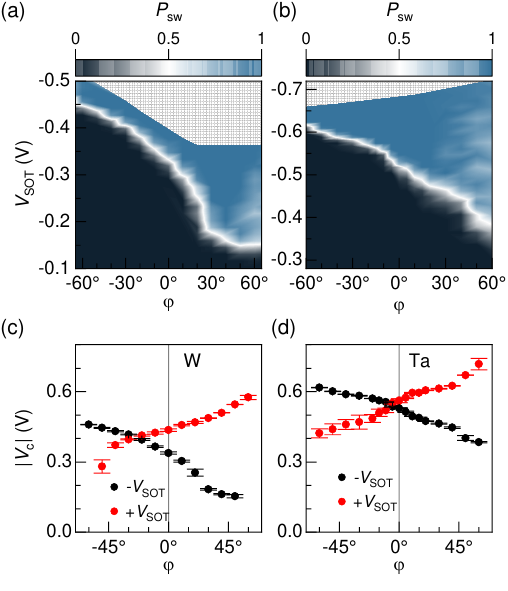}
\subfloat{\label{fig2:a}}%
\subfloat{\label{fig2:b}}%
\subfloat{\label{fig2:c}}%
\subfloat{\label{fig2:d}}%
\caption{\label{fig2:rotB} Switching probability and critical voltage $V_\textbf{c}$ as a function of the angle between $j_{\text{SOT}}$ and $B_{\text{ext}}$. The 2D diagrams in (a,b) show the under-critical (black) and over-critical (blue) conditions for switching to the up state induced by $t_\text{p} = 10$-ns-long current pulses for different orientations of $B_{\text{ext}} = 40$\,mT. The hashed areas indicate the range of parameters which were not investigated in the study. (c,d) Show the corresponding $V_\textbf{c}$ for both pulse polarities. The results for the W and Ta samples are shown in (a,c) and (b,d), respectively.}
\end{figure}

Figure\,\ref{fig2:rotB} summarizes the result of switching by 10-ns-long pulses in a field $B_{\text{ext}} = 40$\,mT applied at different angles $\varphi$ relative to the current.
In both samples, the magnitude of the switching voltage decreases (increases) monotonously with increasing $\varphi$ for negative (positive) SOT pulses, which -- as we discuss further on -- is a manifestation of the superposition of the $y$ component of $B_\text{ext}$ with the effective field of the FLT ($B_\text{FL}$). Figure\,\ref{fig2:a} and \ref{fig2:b} show $P_\text{sw}$ of the up-to-down switching for different SOT pulse and field configurations.
As the critical voltage $V_\text{c} = V_\text{SOT}$($P_\text{sw} = 0.5$) decreases in absolute value for $\varphi > 0$, we observe a broadening of the transition from below- to over-critical voltage, as well as an increased occurrence of switching errors, which becomes significant above $\varphi \approx 30^\circ$. This reduced reliability is a result of the decrease of the longitudinal $B_x$ component required to break the symmetry of the DLT and, possibly, of the FLT-induced precessional dynamics supported by increasing $B_y$.
On the other hand, for $\varphi < 0$, the increase of $|V_\text{c}|$ is not related to deterioration of the switching reliability for over-critical $V_\text{SOT}$. This indicates that the $y$ component of $B_\text{ext}$ can suppress the effect of the FLT even when $V_\text{SOT}$ is large. Moreover, the increase of $|V_\text{c}|$ is accompanied by the narrowing of the transition region (toward negative $\varphi$). This in line with the increase of the switching barrier height, which results in a sharper transition for thermally-activated switching \cite{Lee2014}.

Figures\,\ref{fig2:c} and \ref{fig2:d} compare the absolute values of the critical voltages extracted from the diagrams (black) to those corresponding to the other switching direction (red). Except for small variations, the overall trend of $V_\text{c}$ with $\varphi$ is similar in both samples.
Notably, the trends for the two pulse polarities have opposite slope; the shift of the apparent crossing point to the left of $\varphi=0^\circ$, which can be seen in both panels, is attributed to small differences in the energy landscape of the two reversal directions. These can arise due to structural non-uniformity in the free layer or, more likely, to $B_\text{SAF}$, which makes the switching to the down state (by negative $V_\text{SOT}$) generally more efficient.
The presence of the less reliable switching regions (associated with low $|V_\text{c}|$) defines a finite angular section of $\varphi$ close to $0^\circ$ that can be exploited for tuning the conditions for deterministic bipolar switching (see Sec.\,\ref{sec:outlook}). 
The results of switching by 10 and 1-ns-long pulses are compared in the Supplementary note\,\ref{SI:1ns-pulse}
We observe a similar angular dependence of $V_\text{c}$ for long and short pulses, even though switching by short pulses requires a larger voltage, as expected. The conclusions of the above paragraph are thus independent of the pulse width at least down to 1-ns-long pulses.

To understand the similarity between the W and Ta samples, we note that both are negative spin Hall angle materials and have SOT of equal sign. A positive SOT pulse induces $B_\text{FL}$ pointing along $-y$, as schematized in Fig.\,\ref{fig1:c}, and the Oersted field $B_\text{Oe}$ pointing also along $-y$. From symmetry considerations, the $y$ component of the external field subtracts from (adds up to) $B_\text{FL}$ induced by positive (negative) $V_\text{SOT}$ when $\varphi > 0$; the opposite occurs when $\varphi < 0$. When the two fields add together, the switching is favored, in agreement with previous results obtained in Ta-based samples \cite{Lee2018a,Fan2019}. Ta is known to induce strong FLT, with $\beta$ ranging from 0.7 to over 4 \cite{Garello2013,Kim2013,Avci2014,Qiu2014,Torrejon2014,Ou2016,Lee2018a}, whereas $\beta$ is typically less than 0.4 in the case of W \cite{Garello2018,Sethu2021}. Thus, it may seem surprising that $V_\text{c}$ changes by a similar amount with $\varphi$ in Figs.\,\ref{fig2:c} and \ref{fig2:d}. However, the Ta samples used in this study provide only moderate FLT with $\beta = 1.05\pm 0.08$, and the W samples $\beta = 0.3\pm 0.07$. Moreover, changing $\varphi$ does not affect only $B_\text{FL}$, but also $B_\text{DL}$ that is three times weaker in Ta than in W, which partially compensates for the larger $\beta$ in Ta.

\begin{figure}
\includegraphics[width=85mm]{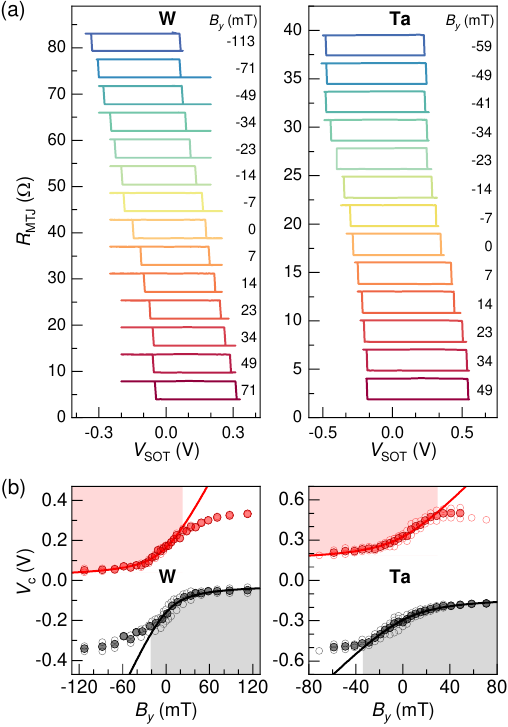}
\subfloat{\label{fig3:a}}%
\subfloat{\label{fig3:b}}%
\caption{\label{fig3:FLT} (a) Switching loops of W- and Ta-based MTJ for different $B_y$ at constant $|B_x| = 40$\,mT. The MTJ resistance is measured as a function of $V_\text{SOT}$ after applying a current pulse with $t_\text{p} = 1$\,ms. The measurements are offset along the $y$ axis for clarity. (b) Critical switching voltage as a function of $B_y$. The full symbols show the average $V_\text{c}$ for positive and negative $B_x$ (open symbols). The lines are fits to the data (see text) in the range marked by the shaded areas.}
\end{figure}

To exclude the effect of insufficient amount of symmetry-breaking field $B_x$, we also study the switching when only the transverse component of the external field $B_y$ is varied and $|B_x|$ is fixed.
For each $B_y$, we send a train of 1-ms pulses of increasing amplitude and determine the switching threshold. Figure\,\ref{fig3:a} compares the switching loops for W and Ta samples measured in $B_x$ = 40\,mT. The extracted thresholds are plotted in Fig.\,\ref{fig3:b}. In both cases, we observe that $V_\text{c}$ changes the most when $B_y$ is close to zero, and then saturates in either field direction. This variation can be understood as the effect of the effective transverse field $B_\text{y}+B_\text{FL}+B_\text{Oe}$ on the switching and confirms that in both systems, $B_\text{FL}$ points along $-y$ when $V_\text{SOT}$ is positive (same sign as in Fig.\,\ref{fig2:rotB}). This trend is opposite to the one observed in Co/Pt/AlO$_\text{x}$ dots, as expected because of the opposite sign of the SOT in Pt with respect to W and Ta \cite{Baumgartner2017}.

Measuring $V_\text{c}$ as a function of $B_y$ further allows for estimating $B_\text{FL}$ at the device level. We start from the analytical formula for the switching threshold $j_{\text{c}}^\perp$ in a transverse field obtained from the Landau-Lifshitz-Gilbert (LLG) equation \cite{Taniguchi2015}, which for $\beta \neq 0$ gives
\begin{equation} \label{eq1:ms}
\begin{split}
j_{\text{c}}^\perp &= \dfrac{2 e M_\textbf{s}t}{\hbar \xi_\text{DL} \beta(2+\alpha\beta)}\cdot\\
&\cdot\left[ (1 + \alpha\beta)B_y \pm \sqrt{2\alpha\beta (2 + \alpha\beta)B_\text{k}^2 + B_y^2}\right].
\end{split}
\end{equation}
Here $M_\text{s}t$ is the unit surface magnetization of the free layer, $\alpha$ is the damping constant, and $B_\text{k}$ is the effective anisotropy field. Theoretically, Eq.\,\eqref{eq1:ms} is valid for $|B_y|$ up to $B_\text{k}$. In the experiment, however, $B_x$ alone can promote the switching, even in the absence of $B_y$. Therefore, the experimental $V_\text{c}$ deviates from the model for $|B_y| \gtrsim |B_x|$ when the sign of $B_y$ is such that it hinders switching. The range of validity of the model is thus reduced to $B_y \leq B_x$ ($B_y \geq -B_x$) for positive (negative) SOT current, corresponding to the red (gray) shaded areas in Fig.\,\ref{fig3:b}.

We fit the data in the shaded regions of Fig.\,\ref{fig3:b} by taking $V_\text{c} = RA_\text{HM}j_{\text{c}}^\perp + V_\text{c0}$, where $A_\text{HM}$ is the cross-section of the HM layer and $V_\text{c0}$ is an offset that takes into account the effect of the constant $B_x$, which is not included in Eq.\,\eqref{eq1:ms}. We take $M_\text{s} = 1.05$\,MA/m, $B_\text{k} = 0.2$\,T (0.27\,T), and $\xi_\text{DL} = -0.325 (-0.108)$ for W (Ta), as measured in full and simplified MTJ stacks and at the device level \cite{Sethu2021,Krizakova2022jmmm},
and let $\alpha$, $\beta$, and $V_\text{c0}$ vary as fit parameters. For simplicity, we neglect the Oersted field $(B_\text{Oe} \approx \mu_0 j_\text{SOT}t_\text{HM}/2)$, which is more than one order of magnitude smaller than the SOT effective fields $(B_\text{DL,FL} = \xi_\text{DL,FL}\hbar j_\text{SOT}/(2 e M_\text{s}t))$ in both samples.
The fits [solid lines in Fig.\,\ref{fig3:b}] give $\alpha = 0.029\pm 0.005$ for W and $0.007\pm0.001$ for Ta, and $\beta = 0.30\pm0.03$ for W and $1.06\pm0.03$ for Ta. Despite the simplifications made, the values of $\beta$ are in close agreement with those obtained from the harmonic Hall measurements [see Sec.\,\ref{sec:experiment}], namely $0.30\pm0.07$ for W and $1.05\pm0.08$ for Ta. 

This result implies that a rather simple model based on the macrospin approximation of the LLG equation can be used to estimate $\beta$, even though the magnetization reversal is incoherent and proceeds via a more complex dynamics \cite{Baumgartner2017,Grimaldi2020}. Moreover, this type of measurement can supplement the hysteresis loop-shift method in finite $B_x$ \cite{Pai2016} that is commonly used to evaluate $B_\text{DL}$ at the device level, to estimate $B_\text{FL}$ without the need to perform harmonic Hall measurements.

\section{\label{sec:TR}Time-resolved switching}
\vspace{-6pt}

\begin{figure}[!b]
\includegraphics[width=85mm]{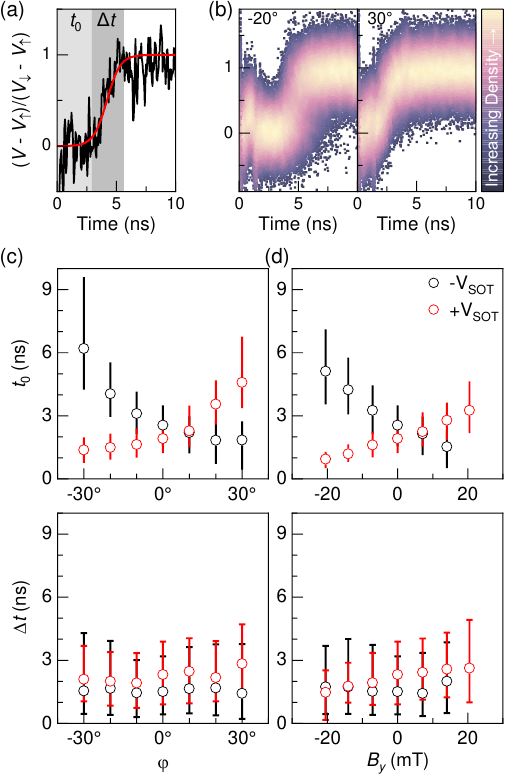}
\subfloat{\label{fig4:a}}%
\subfloat{\label{fig4:b}}%
\subfloat{\label{fig4:c}}%
\subfloat{\label{fig4:d}}%
\caption{\label{fig4:TRss} (a) Representative single-shot switching time trace in a W-based MTJ induced by $V_\text{SOT} = -478$\,mV. An in-plane field $B_\text{ext} = 40$\,mT was applied along $\varphi = -30^\circ$. The delay $t_0$ and duration $\Delta t$ of the reversal (gray shared areas) are found by fitting the data to a sigmoid function (red line). (b) Overlay of 500 time traces acquired at $\varphi = -20^\circ$ (left) and $\varphi = 30^\circ$ (right). (c,d) The activation delay $t_0$ and the transition time $\Delta t$ obtained from fitting the switching traces (c) for different $\varphi$ when $|B_\text{ext}| = 40$\,mT and (d) for different $B_y$ when $B_x = 40$\,mT. The symbols give the median value, the vertical bars give the range of 10--90\% of the events.}
\end{figure}

We next investigate the impact of the FLT and $B_y$ on the switching time scales. Due to the similarity of both types of samples, we only discuss the results for the W-based MTJ and present the results for Ta in Supplementary note\,\ref{SI:TR-Ta}.
We use a constant $V_\text{SOT}$ that ensures reliable switching when 20-ns-long pulses are applied, and probe the magnetization in real time (see Sec.\,\ref{sec:experiment} and Ref.~\onlinecite{Grimaldi2020}). A voltage time trace acquired during each pulse, normalized to the difference between the up and down states [Fig.\,\ref{fig4:a}] represents the perpendicular component of the magnetization of the free layer. In agreement with previous studies \cite{Grimaldi2020,Krizakova2020,Krizakova2021}, each switching event comprises a single transition phase preceded and followed by a quiescent state. By fitting the data to a sigmoid function, we can quantify the activation delay ($t_0$) and the transition time ($\Delta t$) in every measurement. Repeating the acquisition many times in the same conditions provides information about the statistical distribution of the switching times. These characteristic timescales are not accessible when the magnetization state is detected post-pulse, and should not be interchanged with the critical times obtained from the post-pulse switching statistics.

Figure\,\ref{fig4:b} shows an overlay of successful switching events measured in two different current--field configurations, $\varphi = -20^\circ$ (left) and $30^\circ$ (right). For both angles, the traces overlap with a similar dispersion. However, whereas at $-20^\circ$, no reversal starts earlier than 3\,ns after the pulse onset, most reversals at $30^\circ$ are already completed by that time. For any $\varphi$, we did not observe any pauses or intermediate levels in the reversals.

The statistical results for both switching directions as a function of the field orientation (constant $|B_\text{ext}|$) and of the transverse field (constant $B_x$) are summarized in Figs.\,\ref{fig4:c} and \ref{fig4:d}, respectively. The plots of $t_0$ display a strong resemblance to the $V_\text{c}$ dependence in Fig.\,\ref{fig2:rotB}, which corroborates the relation between both quantities \cite{Lee2014,Garello2014}. Accordingly, $t_0$ increases when the transverse field components oppose one another. This shows that $B_y$ (hence also $B_\text{FL}$) directly affects the reversal onset, which is a manifestation of the energy barrier for nucleation of the reversed domain \cite{Miron2010,Grimaldi2020,Krizakova2021}. Due to non-linear scaling of the attempt time with the height of the barrier, the change of $t_0$ is more significant than the change of critical voltage [compare Figs.\,\ref{fig2:rotB} and \ref{fig4:TRss}] in the same range of $B_y$.
On the contrary, the median of $\Delta t$ varies little over the studied range, with only a weak increase with $B_y$ for positive pulses.
This result is consistent with the theory prediction that i) the SOT-driven domain-wall velocity scales with $B_y$, such that $\Delta t$ decreases when $B_\text{FL}$ and $B_y$ are parallel, and ii) the amount of variation increases with the SOT efficiency, current, and Dzyaloshinskii-Moriya interaction (DMI) \cite{Martinez2014, Baumgartner2017}, although the dependence is rather weak unless a strong DMI is involved \cite{Martinez2014}. An increase of the domain wall velocity by 25\% upon changing $B_y$ between $\pm20$\,mT was demonstrated using samples based on Pt, which provides strong DMI \cite{Emori2013a}. However, a small to negligible effect is expected in materials with small DMI, such as Ta or W. In our W samples, the DLT efficiency is about three times higher and the DMI is more than 10 times weaker than in Pt. Therefore, we expect the variation of $\Delta t$ to be less than 10\% in the studied range of $B_y$, consistently with the data in Fig.\,\ref{fig4:TRss}. Similarly, we observe a minor variation of $\Delta t$ in the Ta samples (Supplementary note\,\ref{SI:TR-Ta}).

\vspace{-6pt}
\section{Micromagnetic simulations}
\vspace{-6pt}

In the following, we discuss the effect of FLT and transverse in-plane field in materials with different $\beta$.
To explore SOT switching in a broad range of $\beta$, we have performed micromagnetic simulations using \textsc{MuMax3} \cite{Vansteenkiste2014}. We have simulated the dynamics of a single-layer nanomagnet with 80\,nm in diameter discretized into a $(1.5\times 1.5\times 0.9)$ nm$^3$ mesh. We modeled the free layer using the material parameters of the W/CoFeB system: $M_\text{s} = 1$\,MA/m, $A_\text{ex} = 15$\,pJ/m, $B_\text{k} = 0.25$\,T, $D = 0.2$\,mJ/m$^2$, $\xi_\text{DL} = -0.3$, and $\alpha = 0.1$, and initialize its magnetization along $z$. As in the experiment, we apply a homogeneous magnetic field with different strength and orientation in the plane, and simulate the magnetization reversal induced by spin current to the nanomagnet. The current is supplied by rectangular pulses along $x$ with a 0.1\,ns rising/falling edge. We only discuss here the up-to-down reversal, noting that the simulations are fully deterministic owing to the absence of defects, thermal fluctuations, and SAF field, and thus the down-to-up reversal is the exact opposite of the former.

\begin{figure}
	\includegraphics[width=85mm]{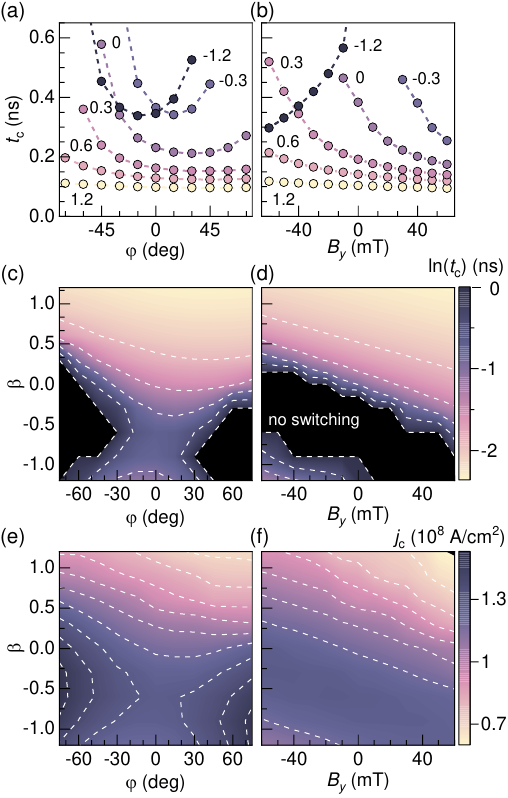}
	\subfloat{\label{fig5:a}}%
	\subfloat{\label{fig5:b}}%
	\subfloat{\label{fig5:c}}%
	\subfloat{\label{fig5:d}}%
	\subfloat{\label{fig5:e}}%
	\subfloat{\label{fig5:f}}%
	\caption{\label{fig5:sims} Results of the zero-temperature micromagnetic simulations. The critical switching time $t_\text{c}$ for different $\beta$ when (a) $|B_\text{ext}| = 40$\,mT and $j_\text{SOT} = -130$\,MA/cm$^2$ and (b) $B_x = 40$\,mT and $j_\text{SOT} = -120$\,MA/cm$^2$. (c,d) 2D diagrams of $\ln(t_\text{c})$ for the parameters used in (a) and (b). (e,f) 2D diagrams of the critical current $j_\text{c}$ as a function of (e) $\beta$ and $\varphi$ ($|B_\text{ext}| = 40$\,mT) and (f) $\beta$ and $B_y$ ($B_x = 40$\,mT).
		The diagrams comprise $13\times 9$ and $11\times 7$ data points, respectively. White dashed lines are isocurves to $\ln(t_\text{c})$ and $j_\text{c}$.}\vspace{-6pt}
\end{figure}

We simulated the switching time traces for different transverse fields (given by $\varphi$ or $B_y$) and $\beta$, and observed that the timing of the reversal, as well as the switching outcome, depend on both the parameters (Supplementary note\,\ref{SI:sim}).
Because not all simulations end in successful reversal, we define $t_\text{c}$ as the time at which the average magnetization of the nanomagnet has undergone the first half of the reversal, i.e., $m_z = 0$.
Figures\,\ref{fig5:a} and \ref{fig5:b} show $t_\text{c}$ for different $|\beta| \leq 1.2$. At each field, we observe that $t_\text{c}$ decreases with $\beta > 0$, because $B_\text{FL}$ in this case supports the tilt of the magnetization induced by $B_\text{DL}$ \cite{Baumgartner2017,Taniguchi2015,Yoon2017}. On the contrary, $B_\text{FL}$ and $B_\text{DL}$ compete when $\beta < 0$, and thus, $t_\text{c}$ first increases with $\beta < 0$ until the effect of $B_\text{FL}$ becomes dominant. Then also the trend of $t_\text{c}$ with transverse field reverses, as visible in Fig.\,\ref{fig5:b}.
Notably, the simulated datasets obtained for $\beta = 0.3$ reproduce the experimental trends, corresponding to $-V_\text{SOT}$ in Figs.\,\ref{fig4:c} and \ref{fig4:d}.
Moreover, one can notice in Fig.\,\ref{fig5:a} that $t_\text{c}$ increases for large $|\varphi|$, regardless of its sign. This effect is a consequence of the reduced $B_x$ and is related to the increase of writing errors observed experimentally in Fig.\,\ref{fig2:rotB}. If $B_x$ is kept constant, on the other hand, $t_\text{c}$ roughly follows an exponential dependence on $B_y$, as shown in Fig.\,\ref{fig5:b}.

Combining the datasets for different $\beta$ produces 2D diagrams of $t_\text{c}$ as a function of $\beta$ and $\varphi$ or $B_y$, as shown in Figs.\,\ref{fig5:c} and \ref{fig5:d}, respectively.
Similarly, we can construct the diagrams for the switching threshold $j_\text{c}$, i.e., the lowest current resulting in the reversal of the magnetization, which is proportional to the variation of the switching energy barrier for different combinations of FLT and external field. Both types of diagrams closely resemble each other, which confirms the relation between $t_\text{c}$ and $j_\text{c}$ \cite{Liu2014a, Lee2014,Garello2014,Raymenants2021a}.
Figure\,\ref{fig5:e} shows that $j_\text{c}$ depends non-linearly on $\varphi$ and $\beta$ with a saddle point at $\varphi = 0^\circ$ and $\beta = -0.8$ obtained for the given simulation conditions. This point marks the conditions, for which the net effect of $B_\text{FL}$ and $B_\text{ext}$ on the magnetization is the smallest. Note that for $\beta$ close to the saddle point, $j_\text{c}$ further rises with $-\cos(\varphi)$, as $B_x$ reduces.
In contrast, Fig.\,\ref{fig5:f} confirms the equivalence of $B_\text{FL}$ and $B_y$ with respect to their impact on $j_\text{c}$. This is visualized by the curves of constant current (white dashed lines), which follow straight lines. Importantly, this further validates the assumptions we made when estimating $\beta$ using Eq.\,\eqref{eq1:ms}. 

\vspace{-6pt}
\section{\label{sec:outlook}Device designs that exploit the FLT}
\vspace{-6pt}

\begin{figure}
	\subfloat{\label{fig6:a}}%
	\subfloat{\label{fig6:b}}%
	\includegraphics[width=85mm]{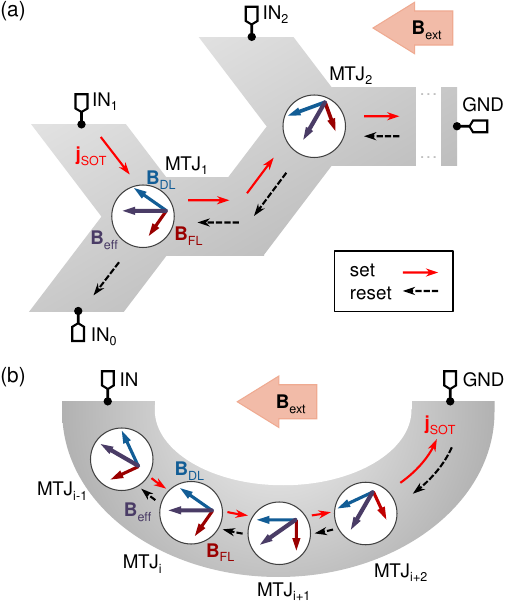}
	\caption{\label{fig6:outlook} Schematics of device concepts that take advantage of the FLT. (a) MTJ array sharing a common SOT track for high-density memory applications. The state of each MTJ can be "set" individually by applying a positive pulse to the respective input electrode, whereas all bits can be "reset" simultaneously by applying a negative pulse to the IN0 electrode. (b) A synaptic weight generator consisting of a series of MTJs sharing a curved SOT track, such that $j_\text{c}($MTJ$_{i}) < j_\text{c}($MTJ$_{i+1}$).}
\end{figure}

Two important challenges in the large-scale adoption of SOT switching for embedded memory applications are the comparatively higher critical current and device footprint compared to spin transfer torque. A useful approach that can tackle both challenges together involves using a common SOT injection path for several MTJ devices and selecting the one to switch by the application of a voltage gate across the MTJ simultaneously with the SOT pulse \cite{Yoda2016, Wu2021}. This scheme, however, requires applying two pulses for writing each bit.
Here, we propose a device design in which the selectivity is intrinsic. This is possible by exploiting the FLT for the MTJ selection [Fig.\,\ref{fig6:outlook}]. Moreover, the proposed geometry allows for reducing the writing energy at the same time.
The schematics in Fig.\,\ref{fig6:a} illustrates this concept on two MTJs placed on a common SOT track and initialized in the down state in the presence of a static magnetic field along $-x$ (provided externally or intrinsic \cite{Garello2019}). The application of a positive "set" pulse on electrode IN$_1$ will induce $j_\text{SOT}$ underneath both MTJs. Due to the geometry of the track, $\varphi < 0^\circ$ for MTJ$_1$, whereas $\varphi > 0^\circ$ for MTJ$_2$. Thus, in line with Figs.\,\ref{fig2:rotB} and \,\ref{fig4:TRss}, the MTJ$_1$ can reverse to the up state at lower bias than the MTJ$_2$. This allows to "set" the MTJ$_1$ without affecting the MTJ$_2$ or any other device on the track. On the contrary, applying a negative pulse to electrode IN$_0$ results in $\varphi > 0^\circ$ underneath all MTJs, which will "reset" all at once to the down state.
From Fig.\,\ref{fig2:rotB}, we can assume reliable switching up to $|\varphi| = 30^\circ$ ($15^\circ$) using W (Ta) under layer and an average difference of $\approx 100$\,mV between the FLT-assisted and hindered thresholds. On top of that, the average writing voltage reduces by 10--20\% compared to $\varphi = 0^\circ$.

Note that another challenge for the SOT switching of perpendicular magnets is the integration of a magnetic field source. In the proposed geometry, we assume a constant $B_\text{ext}$ that can be integrated in devices by a built-in hard mask \cite{Garello2019}, exchange bias \cite{Oh2016} or replaced by other symmetry-breaking mechanisms proposed in the literature \cite{Krizakova2022jmmm}.

A prospective application of MTJs, beyond their use as binary memories, is in computing, as a hardware realization of artificial synapses \cite{Ostwal2019,Akinola2019}. The efficient storing of synaptic weights requires memories with multi-level (preferably analogue) conductance values. A series of MTJs connecting a top and a bottom electrode can serve this purpose \cite{Doevenspeck2021}. It is however crucial that the MTJs can be switched selectively. Figure\,\ref{fig6:b} shows a device that exploits the FLT and the MTJ position on a U-shaped SOT track to enable selective level programming.
As $\varphi$ gradually varies between the pillars along the track, each MTJ will switch at well-defined but different current, since $j_\text{c}($MTJ$_{i}) < j_\text{c}($MTJ$_{i+1}$). To increase the number of weight levels, more pillars can be accommodated on the SOT track. This could be simplified by patterning the track into a "wavy line".
Moreover, the scheme offers potential for very high selectivity, thanks to the possibility to separately optimize the MTJ size and position, the track bending radius, and the FLT strength.

\section{Conclusions}
\vspace{-6pt}

We have studied the influence of the FLT and in-plane magnetic field on the switching of nanoscale magnetic tunnel junctions with a perpendicular free layer and a W or Ta underlayer.
The effective field of the FLT superposes to the component of the in-plane magnetic field transverse to the current. This can be used to reduce or increase the critical switching voltage, the switching reliability, and the activation delay of individual switching events by controlling the magnitude and direction of the external magnetic field. 
We have shown that these effects are significant even in materials with low FLT-toDLT ratio, such as W. On the contrary, the duration of the reversal phase does not considerably change with the transverse field. Together, these results demonstrate that the FLT directly affects the height of the energy barrier for the nucleation of the reversed domain that initiates the switching. Consequently, the FLT strength can be estimated at the device level from measurements of the switching threshold using a simple macrospin model applied to the activation volume, which accounts for the initial domain nucleation phase before domain expansion takes place. We have also performed a systematic micromagnetic study of the critical time and critical current as a function of $\beta$, $\varphi$ and $B_y$. The results of the micromagnetic simulations agree with our experimental findings and allow for predicting the switching behavior in material systems with different FLT strengths. Finally, we proposed two device designs that allow for selectively addressing MTJs sharing the same current-injection track. Selectivity is achieved by varying the alignment of the current, hence of the FLT, relative to the in-plane field, which modifies the critical switching conditions. This approach can be used to create an N-bit memory element with a reduced footprint compared to N separate MTJ devices or a parallel-resistance network with multiple conductance levels allowing for efficient storage and adjustment of synaptic weights.

\begin{acknowledgments}
\vspace{-6pt}
This research was supported by the Swiss National Science Foundation (Grant No.~200020-200465), and imec's Industrial Affiliation Program on MRAM devices. M. H. acknowledges support from the European Union’s Horizon 2020 research and innovation programme under the Marie Sk\l odowska-Curie grant agreement No.~955671.\newline
\end{acknowledgments}

\vspace{20pt}
The datasets presented in this study are available from the corresponding authors upon reasonable request and in the ETH Research Collection with DOI: \href{https://doi.org/10.3929/ethz-b-000569345}{doi.org/10.3929/ethz-b-000569345}

\bibliography{lit}


\onecolumngrid
\newpage

\begin{center}
	\section*{Supplementary material} 
	\vspace{0.5cm}
	\textbf{\large Tailoring the switching efficiency of magnetic tunnel junctions by the fieldlike spin-orbit torque}
	\vspace{0.5cm}
\end{center}

Note 1: SOT switching for different pulse widths

Note 2: Time-resolved measurements in an MTJ with Ta underlayer

Note 3: Simulated magnetization time traces
\vspace{0.5cm}

\renewcommand{\thefigure}{S\arabic{figure}}
\setcounter{figure}{0}

\renewcommand\thesection{{}}
\renewcommand\thesubsection{\arabic{subsection}}

\subsection{\label{SI:1ns-pulse}SOT switching for different pulse widths}

\begin{figure}[hb]
	\includegraphics[scale=1]{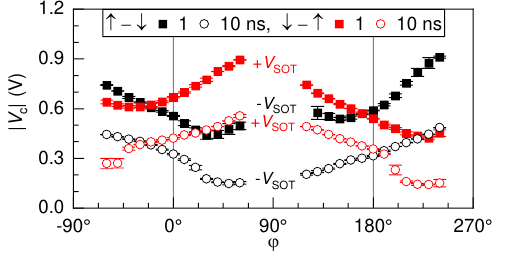}
	\caption{\label{SIfig:W_1-10ns} Dependence of the critical switching voltage $V_\text{c}$ on the direction of $B_\text{ext}$ with respect to $x$ for two different pulse widths. The data are obtained at $|B_\text{ext}| = 40$\,mT using the W-based MTJ.}
\end{figure}

To confirm that the observed variation of the critical switching voltage in an external field $B_\text{ext}$ noncollinear with current is independent of the pulse parameters, we measured the critical switching voltage as a function of $\varphi$ for two different pulse widths. Figure\,\ref{SIfig:W_1-10ns} compares $|V_\text{c}|$ obtained for switching induced by $t_\text{p} = 1$\,ns (full squares) and 10\,ns (open circles), shown in Fig.\,2(c) 
in the main text. Both datasets follow the same trend, including the opposite slope for both pulse polarities, the location of the crossing point on the left of $\varphi = 0^\circ$, and the relative difference of $|V_\text{c}|$ between the reversal supported and hindered by the FLT.

Moreover, Fig.\,\ref{SIfig:W_1-10ns} also shows the switching for the opposite direction of $B_\text{ext}$, i.e., for $\varphi$ close to $180^\circ$. In this case, the longitudinal $B_x$ component has the opposite sign with respect to the switching at $\varphi = 0^\circ$. Thus, the switching polarity is reversed and the positive (negative) SOT pulse induces the up-to-down (down-to-up) reversal. The overall trend however remains the same, in agreement with the explanation by the FLT: the $y$ component of the external field subtracts from (adds up to) $B_\text{FL}$ induced by positive (negative) $V_\text{SOT}$ for $\varphi \in (0^\circ, 180^\circ)$, whereas the opposite occurs for $\varphi \in (180^\circ, 360^\circ)$.
Deterministic switching is not observed for $\varphi$ close to $\pm90^\circ$ due to the absence of symmetry-breaking along the $x$ direction.

\subsection{\label{SI:TR-Ta}Time-resolved measurements in an MTJ with Ta underlayer.}

\begin{figure}[h]
	\includegraphics[scale=1]{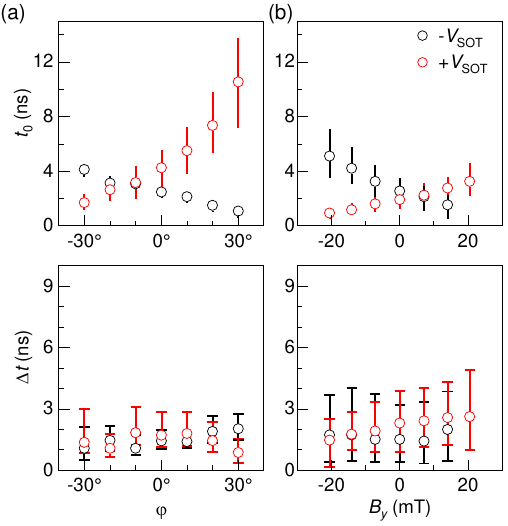}
	\caption{\label{SIfig:TRss_Ta} The activation delay $t_0$ and the transition time $\Delta t$ obtained from fitting the switching traces (a) for different $\varphi$ when $|B_\text{ext}| = 40$\,mT and (b) for different $B_y$ when $B_x = 40$\,mT. All measurements are performed in $B_z = 10$\,mT, which compensates for $B_\text{SAF}$. The symbols give the median value obtained from 550 single-shot measurements, the vertical bars give the range of 10–90\% of these events.
	}
\end{figure}

In the main text, we discuss the switching timescales for an MTJ based on W underlayer. In Fig.\,\ref{SIfig:TRss_Ta}, we show the results measured in a Ta-based MTJ.

Following the same protocol as discussed in the text, we performed single-shot time-resolved switching measurements and extracted the activation delay $t_0$ and the transition time $\Delta t$ from individual switching events. Figure\,\ref{SIfig:TRss_Ta} summarizes the median values of  $t_0$ and $\Delta t$ for different field orientations ($\varphi$) and the transverse fields strengths ($B_y$). Notably, the main results are similar in both types of samples. We observe that $t_0$ is strongly dependent on the external field, and follows the symmetry as observed using the W-based sample, which has the same sign of the SOT. Similarly, $\Delta t$ remains almost independent of the transverse field, as predicted for Ta \cite{Martinez2014}.

However, we also observed certain differences with respect to the measurements using the W-based sample. Whereas W allows for reliable switching for both polarities using the same pulse amplitude and a relatively broad range of $\varphi$, the switching to the up state is less robust against switching errors when over-critical SOT amplitude is applied in the sample with Ta [see Fig.\,1(e) 
in the main text]. Reliable switching in all trials is important for correct evaluation of the switching times. Therefore, we performed the measurements in a constant external field $B_z = 10$\,mT, which compensated the SAF field and equalized the difference between the up and down state. This enabled to achieve reliable bipolar switching in a range of transverse fields.

\subsection{\label{SI:sim}Simulated magnetization time traces.}

We performed a systematic micromagnetic study of the SOT-induced switching as a function of the FLT strength for $|\beta| \leq 1.2$ and the transverse magnetic field given by $|\varphi| \leq 75^\circ$ and $|B_y| \leq 60$\,mT. Each dynamical simulation allows constructing a time trace that shows the evolution of the $z$ component of the magnetization averaged over the nanomagnet volume, before, during, and after the application of the 1-ns-long SOT pulses.
In the simulations, we selected the lowest SOT current that induced the reversal for most combination of the FLT and magnetic field. From each of these time traces, we read the critical switching time $t_\text{c}$ as the time from the pulse onset in which the time trace first crossed $m_z = 0$.

\begin{figure}[hbt]
	\includegraphics[scale=1]{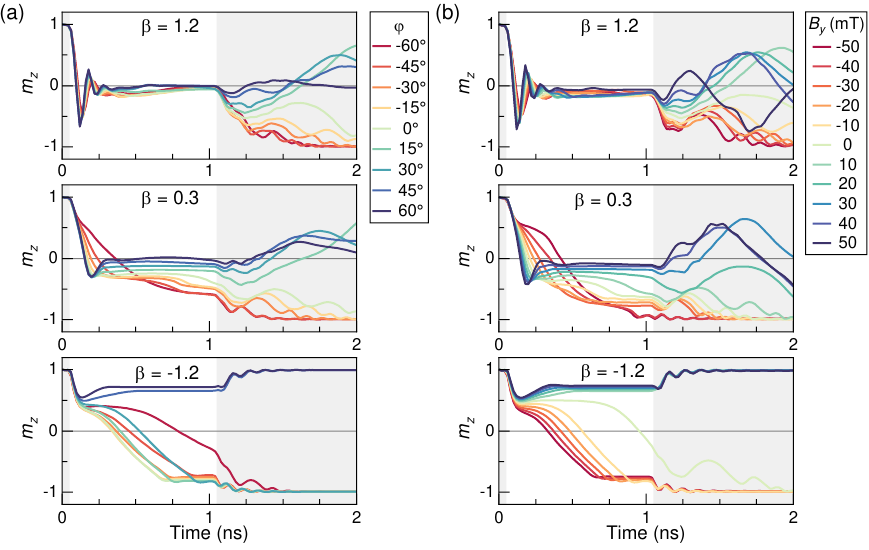}
	\subfloat{\label{SIfig3:a}}%
	\subfloat{\label{SIfig3:b}}%
	\caption{\label{SIfig:sims_traces} Simulated time traces of $m_z$ for three different FLT strengths as a function of (a) $\varphi$ and (b) $B_y$. The time in which the SOT current is "on" ("off") is indicated by the white (gray) background. In (a) $|B_\text{ext}| = 40$\,mT and $j_\text{SOT} = -130$\,MA/cm$^2$, in (b) $B_x = 40$\,mT and $j_\text{SOT} = -120$\,MA/cm$^2$.}
\end{figure}

Figure\,\ref{SIfig:sims_traces} shows representative zero-temperature time traces obtained at $j_\text{SOT} < 0$ and different fields for large positive $\beta$, low positive $\beta$ as in our W-based samples, and large negative $\beta$. The comparison of the three panels shows that the increase of $\beta$ promotes the switching and reduces~$t_\text{c}$.

In these simulations, positive $\beta$ implies that $B_\text{FL}$ shortens the reversal time $t_\text{c}$ for a positive $\varphi$, as it adds to the $y$ component of the external field, in agreement with the measurements reported in Fig.\,4 of the main text. A similar behavior is observed for $\beta = 1.2$ (top panel) and $\beta = 0.3$ (middle panel). However, when $\beta = -1.2$ (bottom panel), $B_\text{FL}$ and the $y$ component of the external field oppose each other at positive $\varphi$, which hinders the reversal. In this configuration, the highest $\varphi$ values correspond to the under-critical conditions, at which the reversal does not initiate.

Beside the effect on $t_\text{c}$, positive $B_\text{FL}$ promotes the magnetization precession. This is revealed in the time traces by the damped oscillations of $m_z$ at the beginning of the current pulse before the equilibrium state with the spin polarization is reached.

After the SOT pulse, the micromagnetic state relaxes to the energy minimum. For large $\varphi$, however, $B_x = B_\text{ext}\cos(\varphi)$ can become too low to efficiently drive a domain wall in the absence of the SOT current, which results in an intermediate state for a prolonged time after the pulse end for positive $\beta$ [top panel in Fig.\,\ref{SIfig3:a}]. Only after a finite time that depends on the damping parameter, the magnetization relaxes to the up or down state. It is, however, less likely to complete the reversal to the final (down) state than in the case of low or negative $\beta$. Experimentally, this would be interpreted as an increased susceptibility to switching errors, as it was observed in Ta- based samples \cite{Lee2018a,Yoon2017}, and attributed to a $B_\text{FL}$-induced reflection of the propagating domain wall from the side of the magnetic layer.

Note that replacing $B_\text{FL}$ by a transverse magnetic field
along the same direction leads to an equivalent outcome. For positive $\beta$, this can be observed in Fig.\,\ref{SIfig:sims_traces} for $\varphi > 0$ or $B_y > 0$.
On the other hand, a transverse magnetic field opposing $B_\text{FL}$ ($\varphi < 0$ or $B_y < 0$) hinders the reversal, which results in longer $t_\text{c}$, but also increases reliability of the switching to the final state. This corresponds well to the experimental results presented in Fig.\,2 
in the main text, where we found that the reliability of the switching to the down state decreases with increasing $\varphi > 0$.

\end{document}